\documentclass[aps,twocolumn,showpacs,nofootinbib]{revtex4-1}

\usepackage[utf8]{inputenc}
\usepackage{graphicx}
\usepackage{amsfonts}
\usepackage{amssymb}
\usepackage{amsmath}
\usepackage[mathscr]{eucal}
\usepackage{setspace}
\usepackage{booktabs}
\usepackage[shellescape,latex]{gmp}
\usepackage{multirow}
\usepackage{color} 
\usepackage[subfigure]{graphfig}
\usepackage{epsfig}
\usepackage{dcolumn}
\usepackage{ulem}

\definecolor{green}{rgb}{0,.5,0}

\definecolor{red}{rgb}{1,0,0}

\def\bea{\begin{eqnarray}}

\def\eea{\end{eqnarray}}

\def\bal#1\eal{\begin{align}#1\end{align}}

\usepackage{hanging}

\def\fir{{\scriptscriptstyle{\text{\rm IR}}}}             
\def\cro{{\scriptscriptstyle{\text{\rm A}}}}              
\def\fa{{\scriptscriptstyle{\text{\rm A}}}}               
\def\efN{\mathscr{N}}                                     
\def\efNm{\efN_\star}                                     
\def\nrN{N}                                               
\def\Lss{{\scriptscriptstyle \text{L}}}                   
\def\Tss{{\scriptscriptstyle \text{T}}}                   

\newcommand*{\GtrApprox}{\smallrel\gtrapprox}

\makeatletter
\newcommand*{\smallrel}[2][.8]{%
\mathrel{\mathpalette{\smallrel@{#1}}{#2}}%
}
\newcommand*{\smallrel@}[3]{%
  \sbox0{$#2\vcenter{}$}%
  \dimen@=\ht0 %
  \raise\dimen@\hbox{%
    \scalebox{#1}{%
      \raise-\dimen@\hbox{$#2#3\m@th$}%
    }%
  }%
}
\makeatother

\immediate\write18{texcount -inc -incbib 
-sum borra.tex > /tmp/wordcount.tex}

\begin{document}

\title{\vspace{1.0in}
Separation of Infrared and Bulk in Thermal QCD 
}
\author{{Xiao-Lan Meng$^{1,2}$, Peng Sun$^{3}$, Andrei Alexandru$^{4}$, Ivan Horv\'ath$^{5,6}$, Keh-Fei Liu$^{6,7}$, Gen Wang$^{8}$, and Yi-Bo Yang$^{1,2,9,10}$}
\vspace*{-0.5cm}
\begin{center}
\large{
\vspace*{0.4cm}
\includegraphics[scale=0.20]{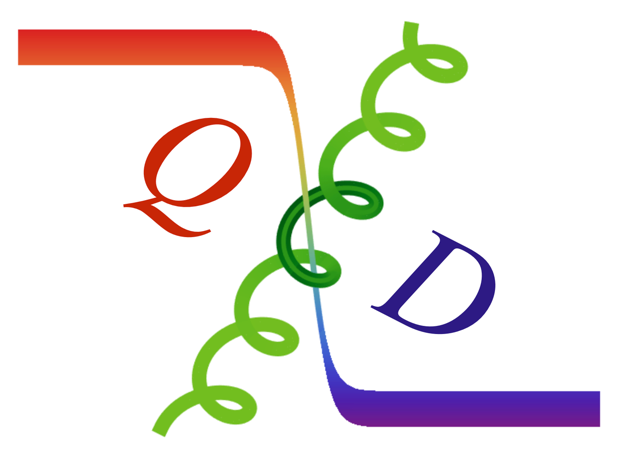}
\includegraphics[scale=0.20]{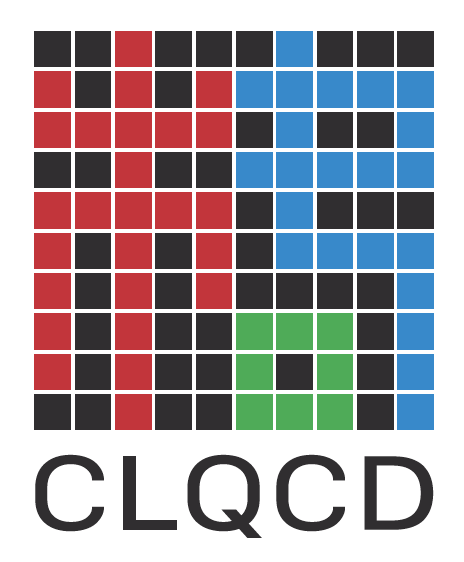}\\
\vspace*{0.4cm}
($\chi$QCD \& CLQCD Collaboration)
}
\end{center}
}
\affiliation{
$^{1}$\mbox{CAS Key Laboratory of Theoretical Physics, Institute of Theoretical Physics, Chinese Academy of Sciences, Beijing 100190, China}\\
$^{2}$\mbox{University of Chinese Academy of Sciences, School of Physical Sciences, Beijing 100049, China}\\
$^{3}$\mbox{Institute of Modern Physics, Chinese Academy of Sciences, Lanzhou, 730000, China}\\
$^{4}$\mbox{The George Washington University, Washington, DC 20052, USA}\\
$^{5}$\mbox{Nuclear Physics Institute CAS, 25068 Rez (Prague), Czech Republic}\\
$^{6}$\mbox{University of Kentucky, Lexington, KY 40506, USA}\\
$^{7}$\mbox{Nuclear Science Division, Lawrence Berkeley National Laboratory, Berkeley, California 94720, USA}
$^{8}$\mbox{Aix-Marseille Université, Université de Toulon, CNRS, CPT, Marseille, France}\\
$^{9}$\mbox{School of Fundamental Physics and Mathematical Sciences, Hangzhou Institute for Advanced Study, UCAS, Hangzhou 310024, China}\\
$^{10}$\mbox{International Centre for Theoretical Physics Asia-Pacific, Beijing/Hangzhou, China}\\
}

\begin{abstract}
A new thermal regime of QCD, featuring decoupled scale-invariant infrared glue, has been
proposed to exist both in pure-glue (N$_f$=0) and ``real-world" (N$_f$=2+1 at physical
quark masses) QCD. In this {\it IR phase}, elementary degrees of freedom flood
the infrared, forming a distinct component independent from the bulk. 
This behavior necessitates non-analyticities in the theory. In pure-glue QCD, 
such non-analyticities have been shown to arise via Anderson-like mobility edges in Dirac spectra 
($\lambda_\fir \!=\! 0$, $\pm \lambda_\text{A} \!\neq\! 0$), as manifested in 
the dimension function $d_\fir(\lambda)$. Here, we present the first evidence,
based on lattice QCD calculation at $a$=0.105 fm,
that this mechanism is also at work in real-world QCD, thus supporting the existence 
of the proposed IR regime in nature. An important aspect of our results is that, while at 
$T\!=\!234\,$MeV we find a dimensional jump between zero modes and lowest near-zero 
modes very close to unity ($d_\fir\!=\!3$ to $d_\fir \!\simeq\! 2$), similar to the IR phase of pure-glue QCD, at $T\!=\!187\,$MeV we observe a continuous 
$\lambda$-dependence. This suggests that thermal states just {\it above} the chiral 
crossover are non-analytically (in $T$) connected to thermal state at $T\!=\!234\,$MeV, 
supporting the key original proposition that the transition into the IR regime occurs at a temperature strictly above the chiral crossover.
\end{abstract}

\maketitle

\textit{1. Introduction}:
Starting with the pre-QCD times of Hagedorn~\cite{Hagedorn:1965st,Hagedorn:1971mc} and early 
lattice QCD calculations in the pure-glue setting~\cite{McLerran:1980pk,Kuti:1980gh,Engels:1980ty}, 
the question of thermal transition in strongly-interacting matter has become one of the highly 
researched topics in nuclear and particle physics. Apart from well-motivated need to understand 
strong interactions, the interest in the issue was fueled, to a large extent, by the potential 
significance of its resolution to the physics of early universe.

Hagedorn in fact set the basic scenario, wherein the thermal transformation process in strong 
interactions boils down to a single ``instability temperature" which, nowadays in QCD, 
is commonly referred as the critical temperature ($T_c$). Due to the non-perturbative nature 
of the problem, lattice QCD became the workhorse for investigations in this area. Advances in 
lattice QCD techniques and computational resources led to a major conclusion, namely that true 
phase transition does not occur in ``real-world" QCD. Rather, {analytic crossovers take 
place} in the temperature range 150-200 MeV, with $T_c \approx 155$~MeV for the case of chiral 
crossover~\cite{Aoki:2006we}.

Transitionless outlook meant a setback to QCD's role in cosmology, but an important 
new twist appeared around the same time. Experiments at RHIC and LHC concluded that 
the state of strongly interacting matter with properties akin to near-perfect fluid 
exists in certain range 
of temperatures~\cite{Arsene:2004fa,Back:2004je,Adams:2005dq,Adcox:2004mh,Muller:2012zq}. 
Among other things, this invites questions about how can such an exotic state of matter 
arise without a true phase transition. 

To this end, some of us presented evidence of an unusual change in QCD Dirac spectra 
at temperatures well above the chiral crossover~\cite{Alexandru:2019gdm}: the anomalous 
accumulation of infrared (IR) Dirac modes, first seen in high-$T$ phase of pure-glue 
QCD~\cite{Edwards:1999zm} and shown to persist into the continuum and infinite-volume 
limits~\cite{Alexandru:2015fxa}, dramatically increases and starts to exhibit signs of 
scale-invariant behavior. This sharp change was found in both the pure-glue and 
real-world QCD ($N_f$ \!=\! 2+1 at physical masses)~\cite{Alexandru:2019gdm}. It was 
proposed that, at the associated temperature $T_\fir$, thermal state of strong 
interactions reconfigures by forming two independent components separated by new scale 
$\Lambda_{\fir}(T) \lesssim T$: the bulk governing distances $\ell \!<\! 1/\Lambda_\fir$ 
and the IR part describing $\ell \!>\! 1/\Lambda_\fir$~\cite{Alexandru:2019gdm}.
In pure-glue case, $T_\fir$ coincides with $T_c$ of Polyakov-line phase transition. 
In real-world QCD, it was also proposed to be a true phase transformation occurring 
somewhere in the range $200 \!<\! T_\fir \!<\! 250$~MeV~\cite{Alexandru:2019gdm}. 
{Its presence may clarify the physics of near-perfect fluid and enhance the role of 
QCD in cosmology.} The associated IR physics only becomes important at volumes unusually 
large by normal QCD standards. In fact, obtaining insights into this emerging IR world 
via numerical simulations requires somewhat unusual 
techniques~\cite{Alexandru:2021pap,Alexandru:2021xoi}.

{The existence of truly distinct high-temperature phase in “real-world” 
QCD~\cite{Alexandru:2019gdm} would naturally be very consequential for understanding 
of strong interactions and for classifying the phases of matter. It is thus very 
important to make the reality of such a basic new feature in thermal QCD as clear 
as possible. The argument of such type turns out to be the most significant output 
of the present study. Its rationale is as follows. Given the existence of an analytic 
chiral crossover at $T_c \!=\! 155\,$MeV, and that a distinct phase transition is 
proposed to occur at $T_\fir \gtrapprox 200\,$MeV, some property of thermal states 
in the intermediate regime ($T_c < T < T_\fir$) should distinguish them non-analytically 
from states in IR phase ($T \!>\! T_\fir$). To that end, we accordingly selected 
temperatures $T \!=\! 187\,$MeV,$\,234\,$MeV, and computed spatial IR dimensions 
of low-lying overlap-Dirac modes which were previously proposed to be among signatures 
playing this role~\cite{Alexandru:2021pap,Alexandru:2021xoi}. Our analysis revealed 
\begin{align}
    \hskip -0.20em 187\, \mathrm{MeV:} \quad\,  d_\fir(0) \!=\! 2.96(22) 
    \quad\;   d_\fir(0^+) \!=\! 3.03(31) \;\;
    \label{eq:936}	\\          
   \hskip -0.03em 234\, \mathrm{MeV:} \quad\,  d_\fir(0) \!=\! 2.98(09) 
   \quad\;   d_\fir(0^+) \!=\! 2.03(16) \;\;
    \label{eq:934}	
\end{align}
at identical UV cutoffs $a \!=\! 0.105 \,$fm. Here $d_\fir(0)$ is the dimension 
of exact zero-modes and $d_\fir(0^+)$ that of lowest non-zero modes. While 
the former doesn't change across our temperature jump, the latter does. In fact, 
the above precisely follows the behavior in $N_f \!=\! 0$ QCD across 
the first-order transition at $T_c \!=\! T_\fir$~\cite{Alexandru:2021pap}. 
Note that $d_\fir$ is the dimension of space available to a quark, 
and {\it any departure} from the  
{constant extended-state value $d_\fir \!=\! 3$, observed at low 
temperatures (see e.g. $T=0$ and $T=187\,$MeV cases at low $\lambda$ 
in Fig.~\ref{fig:d_IR_lambda}),} is expected to generate non-analyticity 
in temperature dependence. The involvement of another integer (topological) 
dimension $d_\fir \!=\! 2$, suggested by our results, is an additional feature 
that may be inherent to IR phase transitions. Such topology aspects may have 
close ties to critical Anderson 
dynamics~\cite{Horvath:2021zjk,Horvath:2022lbj,Horvath:2022klk} 
which is entirely unexpected.}

{To give the above a fuller context, we review in Sec.2 the concept
of IR phase as it developed so far, and describe in Sec.~3 our chief analysis 
tool, namely the effective IR dimension of Dirac modes. We then specify our 
lattice setup and describe our results in Secs. 4, 5 and 6. 
We conclude and further discuss in Sec.~7.}

{\textit{2. IR Phase}:
The notion of new phase in thermal QCD (more generally in SU(N) theories 
with fundamental quarks), reflected in the anomalous accumulation of IR 
Dirac modes, arose in Ref.~\cite{Alexandru:2015fxa}. This work showed 
that the effect is not an UV or IR artifact, and associated it with 
the deconfinement of those QCD degrees of freedom that separate in this
way from the rest. Based on extensive new data, 
Ref.~\cite{Alexandru:2019gdm} made a key finding that the first-order 
phase transition in $N_f \!=\! 0$ QCD marks the onset of IR power-law 
divergence in Dirac spectral density $\rho(\lambda)$ (negative near-pure 
power $p \!\GtrApprox\! -1$), and that similar change occurs 
in real-world $N_f \!=\! 2\!+\!1$ QCD at temperature $T_\fir$ distinctly 
above the chiral crossover. The ``anomalous accumulation" of IR 
modes was thus identified with $p \!<\!0$. Such regime was immediately 
associated~\cite{Alexandru:2019gdm,Alexandru:2021pap} with IR-bulk 
separation (decoupling), scale-invariant behavior of IR glue including 
infinite glue screening lengths, and non-analyticities in Dirac spectrum 
leading to non-analytic T-dependence of observables at $T_\fir$. 
These properties are built into the {\it metal-to-critical} picture 
of IR phase transition~\cite{Alexandru:2021xoi} based on the existence 
of two Anderson-like mobility edges: the previously studied 
$\lambda_\fa \!>\! 0$~\cite{GarciaGarcia:2005vj, GarciaGarcia:2006gr,
Kovacs:2010wx, Giordano:2013taa,Ujfalusi:2015nha}, and the new 
$\lambda_\fir \!=\! 0$. 
}

{Effort to organize the accumulated knowledge naturally led to defining 
the phases based on $p$ in 
$\rho(\lambda) \!\propto\! \lambda^p$ for $\lambda \!\to\! 0$,
namely~\cite{Alexandru:2019gdm} 
\begin{equation}
     \text{phase}  \;=\;
     \begin{cases} 
        \;\;\text{B}  $\quad$ \text{(IR-broken)}     &     \text{if} \quad  p=0  \\[2pt]
        \;\;\text{IR} $\;\;$ \text{(IR-symmetric)}   &     \text{if} \quad  p<0  \\[2pt]
        \;\;\text{UV} $\;$ \text{(IR-trivial)}       &     \text{if} \quad  p>0        
     \end{cases}
     \quad
    \label{eq:964}	     
\end{equation} 
where the adjectives relate to scale invariance of glue. In case of real-world 
thermal QCD, B is the usual confined phase at zero and low temperatures. 
As clarified already in the original work~\cite{Alexandru:2019gdm}, 
$\rho(\lambda)$ IR-diverges even in B, but it is the logarithmic divergence 
predicted by chiral perturbation theory~\cite{Osborn:1998qb}, 
and thus $p\!=\!0$. While this virtually 
guarantees that $T_\fir$ defined by \eqref{eq:964} exists in real-world QCD, 
its precise determination is laborious since it is difficult to distinguish 
a strong logarithm and a small power.\footnote{{It should also be pointed 
out that there is no solid evidence as yet for the existence of thermal UV phase 
which corresponds to fully perturbative quark-gluon plasma~\cite{Alexandru:2019gdm}}.} 
Given that, we mainly focus here on the inner non-analyticity aspect of IR phase 
which, in turn, is most directly tied to IR-bulk 
separation~\cite{Alexandru:2019gdm,Alexandru:2021pap}. 
In fact, all aspects of IR phase listed in the previous paragraph are closely 
interconnected, and in $N_f\!=\!0$ QCD materialize at single temperature 
$T_\fir \!=\! T_c$. In real-world QCD the existence of a single transition 
temperature is conjectural at present due to ensuing uncertainties. However, 
the transition associated with {\em any one} of these features would in itself 
signal a new phase transformation in QCD.}

{\textit{3. Effective Dimensions}:
Predicted non-analyticities of Dirac spectra in IR phase~\cite{Alexandru:2019gdm} 
are realized in IR-dimension function $d_\fir(\lambda)$ of 
modes~\cite{Alexandru:2021pap,Horvath:2022ewv}. This construct conveys details 
on the possibility of space effectively collapsing into a lower dimension for 
any part of the spectrum, which was indeed observed~\cite{Alexandru:2021pap}. 
That invokes possible associations with Anderson transitions in condensed-matter 
systems and, indeed, the behavior of $d_\fir(\lambda)$ upon entering the IR phase 
was interpreted as {\it metal-to-critical} transition of 
Anderson-like~type~\cite{Alexandru:2021xoi}.
}

{The general concept of effective spatial dimensions $d_\fir$~\cite{Horvath:2022ewv} 
arose from very recent advances in assigning effective measures to regions endowed with probabilities, 
the so-called effective-number theory~\cite{Horvath:2018aap,Horvath:2018xgx,Horvath:2019qeo}. 
At the technical level, it boils down to replacement of ordinary counting in the definition of 
box-counting dimension for fixed sets by effective counting for probability distributions.}
For Dirac modes in thermal QCD the prescription is as follows. In lattice-regularized 
Euclidean setup the number of 
sites $\nrN(L) \equiv (L/a)^3/(aT)$ (UV cutoff $1/a$, IR cutoff $1/L$, temperature $T$) grows as 
$L^3$ at fixed $a$, conveying that IR dimension of space is $D_\fir \!=\!3$.
But Dirac eigenmode $D \psi(x) \!=\! \lambda \psi(x)$ entails probabilities 
$P \!=\! (p_1,p_2,\ldots,p_\nrN)$, $p_i \!\equiv\! \psi^+ (x_i)\psi(x_i)$, and sites 
have to be counted {\it effectively} in order to quantify the volume $\psi$ actually extends 
into, namely~\cite{Horvath:2018aap} 
\begin{equation}
     \nrN\, \longrightarrow\, \efNm[\psi] \,=\, \efNm[P] \,=\, 
     \sum_{i=1}^\nrN \min\, \{ \nrN p_i, 1 \}    \;\,
     \label{eq:023}              
\end{equation}
The IR scaling of QCD-averaged effective volume at given $\lambda$ then 
determines $d_\fir(\lambda)$ at UV cutoff $a$, namely~\cite{Alexandru:2021pap,Horvath:2022ewv}
\begin{equation}
     \langle \, \efNm \,\rangle_{L,\lambda,a} \,\propto\, L^{d_\fir(\lambda,a)} 
     \quad \text{for} \quad L \to \infty  \quad 
     \label{eq:028}	
\end{equation} 

Using the overlap Dirac operator due to its superior chiral
and topological properties, an unusual $d_\fir(\lambda)$ was found in IR phase of 
pure-glue QCD~\cite{Alexandru:2019gdm}. Indeed, the function is non-analytic at both 
$\lambda_\fir$ and $\lambda_A$, with spectral region of low-$d$ modes between 
them~\cite{Alexandru:2021pap}. Moreover, in contrast to exact zero modes, which 
are $d_\fir \!=\!3$, the lowest near-zero modes $(\lambda \!\to\! 0^+)$ 
are close to other topological value $d_\fir \!=\!2$. Such jump at $\lambda_\fir \!=\!0$ 
is surprising since the proposed origin of anomalous IR mode accumulation is 
the conventional mixing of topological lumps~\cite{Edwards:1999zm,Vig:2021oyt} which, 
in absence of additional (unknown) effects, leads to $d_\fir \!=\! 3$ in both cases.

\textit{4. Numerical Setup and Dirac Spectral Density:}
We lattice-regularize $N_f \!=\! 2+1$ QCD using tadpole-improved clover fermion action 
(1-step stout link smearing with parameter 0.125) and tadpole-improved Symanzik gauge action 
at $a \!=\!0.105\,$fm and $m_{\pi}\simeq 135\,$MeV. 
Ensembles at temperatures $T \!=\! 187$ and $234\,$MeV on numerous spatial volumes 
(up to $L\!=\!10.1\,$fm) were generated by CLQCD collaboration (see Table~\ref{Tab:setup}), 
which allows for calculation of IR dimensions. More detailed ensemble description is given 
in Ref.~\cite{Hu:2023jet}. We note in passing that ensembles with similar quark and gauge actions 
were already used in previous zero-temperature
calculations~\cite{Zhang:2021oja,Liu:2022gxf,Xing:2022ijm,Liu:2023feb,Hu:2023jet}.

\begin{table}[t] 
\vspace{-0.10in}
\centering
\caption{UV cutoff $a$, pion mass $m_{\pi}$, lattice volumes $n_{L}^3\times n_T$ and temperature $T$ of lattice QCD ensembles studied.}
\label{Tab:setup}
\begin{tabular}{cclrc}
\hline
\hline
$a$(fm) & $m_{\pi}$(MeV) & $n_L$ & $n_T$ & $T$(MeV) \\
\hline
 0.105 & 135 & 24/28/32/40/48/64/96 & 8 & 234\\
 0.105&  135 & \quad \quad \ \ \  32/40/48/64 & 10 & 187\\
\hline
\vspace{-0.35in}
\end{tabular}
\end{table}

Lattice glue fields $U$ in these theories will be studied via their effect on the overlap Dirac operator 
$D_\text{ov}[U]$. We construct $D_\text{ov}$ using the original square-root 
prescription~\cite{Neuberger:1997fp} at $\rho \!=\!1.5$ with 1-step HYP smearing of $U$. 
To determine the low-lying eigensystem, we select the chiral sector containing zero mode(s), 
calculate the eigenvectors of $D_{\rm ov}^{\dagger}D_{\rm ov}$ in it using the Arnoldi method, and then 
construct non-zero modes~\cite{Sorensen:1992fk,Lehoucq:1996,Giusti:2002sm,Alexandru:2011sc}. 
Transformation 
$D \equiv D_{\textrm{ov}}/(1-\frac{a}{2\rho}D_{\textrm{ov}})$ \cite{Chiu:1998gp} yields 
purely imaginary eigenvalues ($D \psi_\lambda(x) \!=\! i\lambda \psi_\lambda(x)$) and 
the associated spectral density is $\rho(\lambda) = T \sum_{i} \delta( \lambda-\lambda_i)/L^3$. 
Further technical details can be found in Appendix~\ref{app:accuracy}.

\begin{figure}[t] 
\centering
\includegraphics[scale=0.5]{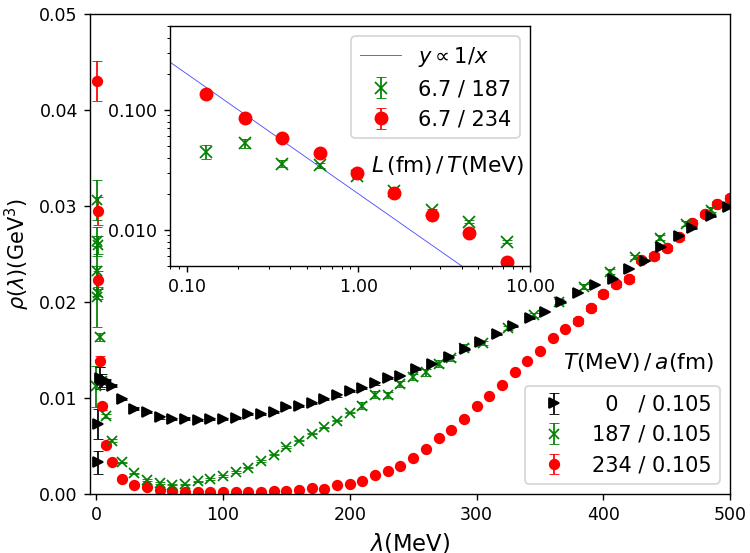}
\caption{Spectral density $\rho(\lambda)$ at $T \! \simeq \! 0\,$MeV (triangles), 
$T \!=\! 187\,$MeV (crosses) and $T \!=\! 234\,$MeV (circles). The cutoffs are 
$a \!=\! 0.105\,$fm and $L\!=\!5.0\,$fm in all cases. At $T \!\simeq\! 0~$MeV, 
$\rho(\lambda)$ was computed using stochastic estimates~\cite{Cossu:2016eqs}.
{The inset shows a log-log plot of density in deep IR. See text for details.}}
\vspace{-0.21in}
\label{fig:densities}
\end{figure}

\begin{figure}[t]
\centering
\includegraphics[scale=0.5]{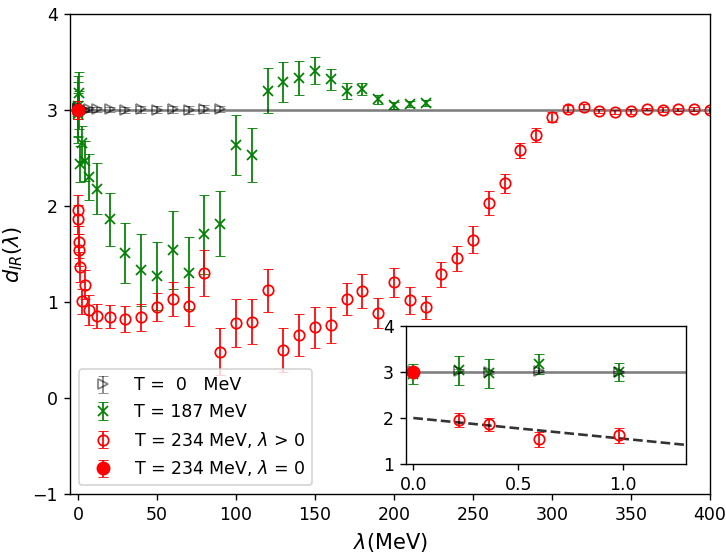}
\caption{Function $d_{\rm IR}(\lambda)$ at $T \!=\!0$ (triangles), $T \!=\!187\,$MeV 
(crosses) and $T \!=\!234\,$MeV (circles). The full circle is the result for exact zero modes
at $T \!=\!234\,$MeV. The inset zooms in on deep infrared with linear $\lambda \to 0^+$ 
extrapolations shown.}
\vspace{-0.2in}
\label{fig:d_IR_lambda}
\end{figure}

Eigenmodes with $\lambda$ up to $\approx \!500$ MeV were computed for ensembles in 
Table~\ref{Tab:setup}. Densities $\rho(\lambda)$ were calculated and renormalized in $\overline{\textrm{MS}}$ 
at 2 GeV, using $Z_m \!=\!Z^{-1}_S \!=\! 0.907(26)$ obtained by interpolating the results at 11 UV 
cutoffs~\cite{He:2022lse}. Our main interest is the temperature range $200 \!<\! T \!<\! 250$~MeV, 
where the system was originally predicted  to reach the IR phase at certain $T_\fir$. 
In Fig.~\ref{fig:densities} we show $\rho(\lambda)$ at $T \!=\!234~$MeV (red circles). 
The striking bimodal structure exhibits features previously associated with IR 
phase~\cite{Alexandru:2019gdm}, including a fully-formed region of mode depletion: the plateau. 
For comparison, we also show densities at $T \!=\! 187\,$MeV (green crosses) and 
$T \!\simeq\! 0~$MeV (black triangles). The three simulations were performed in the same lattice setup 
and at identical parameters except for $n_T \!=\! 8,10,96$ respectively, which controls the temperature.
The changes in the displayed sequence of Dirac spectra are thus exclusively due to the changing level 
of thermal agitation. In fact, the three cases illustrate three thermal regimes of 
Ref.~\cite{Alexandru:2019gdm} in terms of IR-bulk separation.
Indeed, the $T \!\simeq\! 0$ case showcases the domain $0 \!\le\! T \!<\! T_A$ 
($T_A \!\approx\! T_c \!\approx\! 155\,$MeV) where IR and bulk are not meaningfully separable. 
At $T \!=\! 187\,$MeV, the system is inside the  finite transition regime $T_A \!<\! T < T_\fir$ 
where the separation process effectively starts at $T_A$ and becomes complete at $T_\fir$ which 
is the onset of IR phase.  At $T \!=\!234~$MeV, IR and bulk seem fully separated (IR phase),
consistently with the original estimate $T_\fir \in (200,250)\,$MeV~\cite{Alexandru:2019gdm}.

{The inset of Fig.~\ref{fig:densities} shows the comparison of densities at $T\!=\!234$ and 
$187\,$MeV in a log-log plot of deep IR on the largest common volume. A line sloped as 
$\lambda^{-1}$ is also shown to guide the eye. Note that below about $0.5\,$MeV 
(well below {our light-quark mass, which is 3.6(2) MeV~\cite{Hu:2023jet} in the 
renormalization scheme also used for $\lambda$ and $\rho(\lambda)$}), the behavior at 
$T \!=\! 234\,$MeV approximately adheres to the IR rise of density corresponding 
to $p \GtrApprox -1$, while at $T\!=\!187\,$MeV there is a clear tendency to level off. 
This aspect thus also points toward two different kinds of thermal states.
}

{Although the above considerations are quite cogent, they are too qualitative to 
make sharp statements about the situation at hand. We thus turn to the non-analyticity 
aspect of IR phase~\cite{Alexandru:2019gdm, Alexandru:2021pap, Alexandru:2021xoi}, which 
at present uses the IR dimension function $d_\fir(\lambda,T)$ as its basic tool. 
In pure-glue QCD, $d_\fir(\lambda,T)$ becomes non-analytic (discontinuous) at 
$\lambda \!=\! 0$~\cite{Alexandru:2021pap, Alexandru:2021xoi} upon crossing into the IR 
phase. Since IR dimension is a global characteristic of modes, we also expect it to be 
a very robust indicator of phase changes it reflects. Indeed, we will provide evidence 
that the innate difference between $T \!=\! 187$ and $234\,$MeV is cleanly revealed in 
this way.
}

Before proceeding to make this point, let us comment in passing on a different aspect
conveyed by Fig.~\ref{fig:densities}. In particular, it is generally assumed, based on asymptotic 
freedom, that UV behavior of $\rho(\lambda)$ is the same at any temperature $T$. However, it is not 
a priori clear from such arguments how and at what scales this occurs. Remarkably, we see that
$\rho(\lambda,T\!=\! 187\, \text{MeV})$ not only approaches but becomes effectively indistinguishable 
from $\rho(\lambda,T\!=\! 0)$ at unexpectedly small $\lambda\sim$  270\,\text{MeV}. Indeed, the two dependences join
above the scale of temperature (187 MeV) but well before the asymptotic behavior 
($\rho(\lambda) \propto \lambda^3$) sets in.  System at $T\!=\! 234\,$MeV follows 
the same pattern.

\textit{5. IR Dimensions of Dirac Eigenmodes:} 
We now examine in detail whether the unusual $d_\fir(\lambda)$ in IR phase 
of pure-glue QCD~\cite{Alexandru:2021pap} is also manifested in real-world 
QCD at $T \!=\!234\,$MeV and $T\!=\!187\,$MeV. 
To that end, we utilize and extend the techniques of early studies. 
A useful concept is the ``finite-volume" $d_\fir$, namely~\cite{Alexandru:2021pap}
\begin{equation}
  d_\fir(L,s) = \frac{1}{\ln(s)} \ln \frac{\efNm(L)}{\efNm(L/s)} \quad , \quad 
  s > 0
  \label{eq:64}     
\end{equation}  
since then $d_\fir \!=\! \lim_{L \to \infty} d_\fir(L,s)$ independently of $s$.
Estimating the limit from linear extrapolations in $1/L$ works well in Anderson models, 
at least for extended states and at criticality~\cite{Horvath:2021zjk}. Here we utilize 
this, and point out that the procedure is equivalent to direct fitting of $\efNm(L)$ 
to the form $b\, L^{d_\fir} e^{-c/L}$ (see Appendix~\ref{app:fitting}), which is 
technically more convenient.

\underline{$T\!=\!234\,$MeV}.
Using the data on five largest systems to fit, we obtained $d_\fir(\lambda)$ 
shown in Fig.~\ref{fig:d_IR_lambda}. Despite some differences (see below), 
its behavior is strikingly similar to pure-glue case (Fig.~1 of Ref.~\cite{Alexandru:2021pap}).
Important commonality is the discontinuity feature at $\lambda_\fir \!=\! 0$, suggesting 
that exact zero-modes ($d_\fir(0) \!\simeq\! 3$, full red circle) differ from lowest near-zero 
modes ($d_\fir(0^+) \!\simeq\! 2$, red circles) in a robust qualitative manner. This is made 
explicit by the inset of Fig.~\ref{fig:d_IR_lambda} focusing on the very deep IR, and yielding 
$d_\fir(0^+) \!=\! 2.03(16)$ after linear $\lambda \!\rightarrow\! 0^+$ extrapolation. 
Explanation of this (more than $5 \sigma$) difference in terms of the underlying IR glue may 
provide important clues toward the full understanding of IR phase.

Like in pure-glue case, we find a clear low-$d$ plateau, here in the range of about 
$10\!-\!220\,$MeV, roughly coinciding with the region of strongly suppressed 
$\rho(\lambda)$ (Fig.~\ref{fig:densities} vs Fig.~\ref{fig:d_IR_lambda}). Dimensional 
structure of plateaus will be further clarified in forthcoming studies using 
the multidimensionality technique of Ref.~\cite{Horvath:2022klk}. 

The onset of rise toward dimension 3 past $\lambda \!\approx\! 220\,$MeV confirms the viability 
of scenario with mobility edges $\lambda_\fir$ and $\pm\lambda_A$~\cite{Alexandru:2021xoi}. 
However, the discontinuity of $d_\fir(\lambda)$ at presumed $\lambda_A$ is not apparent 
in our data, contrary to both the pure-glue case~\cite{Alexandru:2021pap} and  the situation 
in Anderson models~\cite{Horvath:2021zjk}. Resolution of this difference may provide 
an additional new insight into the IR-phase dynamics. 

\underline{$T \!=\! 187\,$MeV}. Data for all four system sizes were used in fits 
to obtain $d_\fir(\lambda)$ shown in Fig.~\ref{fig:d_IR_lambda}. Here the situation is 
different, most visibly in the approach to deep infrared (see the inset). In particular, 
we find no detectable disconnect between exact zero modes and the deepest near-zero modes. 
In fact, in this case we have $d_\fir(0^+) \!=\!3.03(31)$ after linear 
$\lambda \!\rightarrow\! 0^+$ extrapolation in clear contrast to $T \!=\!234\,$MeV. 

The dip of $d_\fir(\lambda)$ in the range of about 10-90 MeV is a structure qualitatively 
between a low-$d_\fir$ plateau of $T\!=\!234\,$MeV and a constant behavior typical of low 
temperatures. We thus cannot say unambiguously whether this aspect is IR phase-like or not.  
However, the moderate unphysical overshooting of $d_\fir \!=3\!$ above the dip and rather 
large error bars suggest that better statistics and simulation of somewhat larger systems
will provide the soon-accessible answer to this question.

 \underline{$T \!=\! 0$}.
It is instructive in this context to also perform a $d_\fir$ calculation at zero temperature.  
To that end, we used three 2+1 flavor Mobius+Iwasaki+DSDR ensembles from the RBC 
collaboration at $a\!=\!0.194\,$fm and  $n_L=24,32,48$~\cite{RBC:2012cbl,Boyle:2015exm}.
With three sizes available, we performed two-parameter fits of the form $bL^{d_\fir}$ with
results shown in Fig.~\ref{fig:d_IR_lambda} (triangles). Note that the available range 
of $\lambda$ is smaller due to the large number of lattice points at low temperatures.
However, apparently sufficient statistics and lattice sizes yield a stable 
$d_\fir(\lambda) \!\equiv\! 3$, as expected.

\begin{figure}[b]
\vspace{-0.18in}
\centering
\includegraphics[scale=0.46]{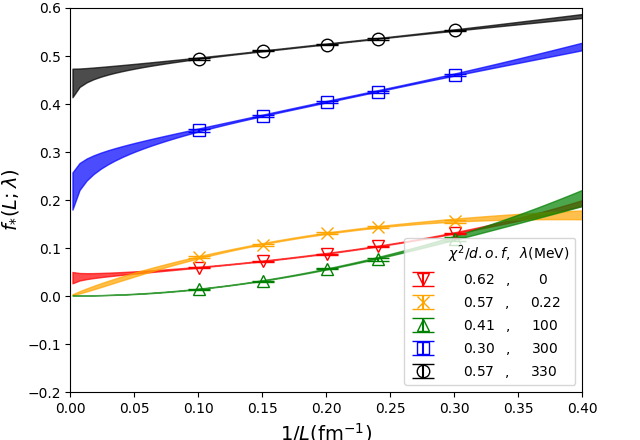}\\
\caption{Function $f_\star(L)$ for various $\lambda$ at $T\!=\!234\,$MeV.} 
\label{fig:f_star_lambda}
\end{figure}

\textit{6. IR Scaling:}
To illustrate the quality of scaling in various $\lambda$-regimes shown in 
Fig.~\ref{fig:d_IR_lambda}, we plot in Fig.~\ref{fig:f_star_lambda} 
the fraction $f_\star$ of volume taken by the effective support of the state, 
namely $f_\star \!\equiv\! \efNm/ \nrN \!=\! \efNm/(n_\Lss^3 n_\Tss^{})$. 
Since $\efNm(L) \!\propto\! L^{d_\fir} e^{-c/L}$ is used to extract $d_\fir$, we 
have $f_\star(L) \!\propto\! L^{d_\fir -3} e^{-c/L}$ and these fits are shown 
in Fig.~\ref{fig:f_star_lambda} for $T\!=\! 234\,$MeV. 
The displayed $\chi^2$/dof for modes in different 
regimes do indeed confirm very good scaling behavior. Note how functions $f_\star(L)$ 
in Fig.~\ref{fig:f_star_lambda} visually separate the bulk modes and near-bulk modes 
from~IR~modes. Indeed, although zero-modes are $d_\fir \!=\!3$, and hence occupy
a finite fraction of volume in thermodynamic limit ($\lim_{L\to\infty} f_\star(L) \!>\!0$), 
its magnitude is much smaller than that of typical bulk modes. At the same time, for 
$d_\fir \!<\! 3$ modes of IR component the effective fraction vanishes 
($\lim_{L\to\infty} f_\star(L)\!=\!0$).

Fig.~\ref{fig:f_star_lambda} reveals that the lowest near-zero modes we studied 
($\lambda \!=\! 0.22\,$MeV, $d_\fir \!<\! 3$) have larger $f_\star$ at studied volumes than those 
of zero modes ($\lambda \!=\! 0$, $d_\fir \!=\!3)$. But given their $d_\fir$, this order has 
to reverse at sufficiently large volume. We can read off the graphs in Fig.~\ref{fig:f_star_lambda} 
that this happens at $L \approx 20\,$fm. Such deep IR thresholds simply do not appear in other QCD 
regimes. Only at larger $L$ will modes at $\lambda \!=\! 0.22\,$MeV become ``sparser" than zero modes.
Note that the qualitative difference between zero and near-zero modes is expressed here by the 
opposite convexity properties of their $f_\star(L,\lambda)$. 

\begin{figure}[t]
\centering
\vspace{0.04in}
\hspace{-0.11in} \includegraphics[scale=0.42]{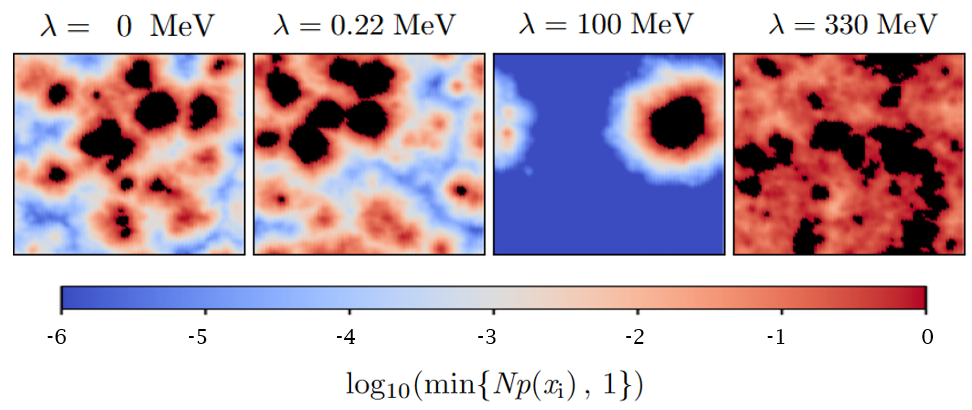}
\vspace{-0.19in}
\caption{Typical color-coded $\log_{10}(\min\, \{\nrN p(x_i), 1 \})$ in a 2d plane 
containing $x_i$ with maximal probability. Modes in different $\lambda$-regimes 
are shown at $T\!=\!234\,$MeV and $L\!=\!10.1\,$fm.}
\label{fig:dist_lambda}
\vspace{-0.21in}
\end{figure}

Finally, we wish to gain some visual insight into the spatial geometry of modes. In
definition~\eqref{eq:023} of $\efNm$, uniform probability $p_u \!=\! 1/\nrN$ 
enters as a reference value: points $x_i$ with 
$p(x_i) \!=\!\psi_\lambda^\dagger(x_i) \psi_\lambda(x_i) \!\ge\! p_u$ are guaranteed 
to be in effective support, and we refer to them as ``core". 
We wish to set up a sea-level representation that visualizes it sharply. 
Plotting $\min \{ \nrN p(x_i), 1 \}$, namely the contribution of $x_i$ to 
effective count, accomplishes that. In Fig.~\ref{fig:dist_lambda} we color-code this input 
(on a logarithmic scale) and show its typical behavior in a plane containing the global 
probability maximum. The black regions mark the core. The panels represent different 
$\lambda$-regimes on the same glue background. The bulk mode at $\lambda \!=\!330\,$MeV (right) 
resembles modes at low temperatures in that its core spreads out contiguously over large 
distances and its granularity (composition from distinct lumps) is not very obvious. 
To the contrary, the plateau mode ($\lambda \!=\!100\,$MeV) is usually dominated by a well-formed 
lump as shown. The near-zeromodes ($\lambda \!=\! 0.22\,$MeV) maintain the granularity, but 
involve multiple lumpy features forming a larger spatial structure. The zero-modes~(left) at this 
volume are in fact quite similar but, due to $d_\fir \!=\! 3$ vs $d_\fir \!=\! 2$ difference, 
will become infinitely more ``space-filling" in thermodynamic~limit.

Additional results are described in Appendices~\ref{app:fitting},\ref{app:Visualization}.

\textit{7. Summary and Discussion:}
Remarkable property of QCD IR phase~\cite{Alexandru:2019gdm} is that it requires the presence 
of inner non-analyticities not only at the transition point $T_\fir$, but at any temperature
within the phase. It was proposed and verified~\cite{Alexandru:2021pap,Alexandru:2021xoi} 
that in pure-glue QCD the system arranges for this by reconfiguring itself into independent 
parts (IR \& bulk), sharply separated in Dirac spectrum. The needed non-analyticity enters 
via Anderson-like mobility edges $\lambda_\fir \!=\!0$ and $\pm \lambda_\text{A} \!\neq\! 0$, 
encoded by the dimension function $d_\fir(\lambda)$. In this work we presented results 
suggesting that key elements of this scenario also materialize in ``real-world" QCD at 
$T\!=\!234\,$MeV. 
Thus, in certain regards, thermal state in IR phase of strong interactions resembles 
the Tisza-Landau two-fluid model of liquid helium. The proposed 2-component nature of thermal 
state may in fact be the most essential attribute of IR phase.

The above is based on our results from simulations of $N_f\!=\! 2 \!+\! 1$ QCD at 
physical quark masses, UV cutoff $a\!=\!0.105\,$fm and temperatures $T\!=\!234$ and 
$187\,$MeV. The former temperature is in the range where the transition temperature 
$T_\fir$ was originally estimated ($200 \!<\! T_\fir \!<\! 250\,$MeV) and our results 
suggest that $T \!=\! 234\,$MeV is past that onset. Indeed, here the computed IR-dimension
function $d_\fir(\lambda)$ exhibits the same broad features as that of pure-glue QCD 
in IR phase. Particularly striking is the clear presence of the identical tell-tale 
discontinuity at $\lambda \!=\! 0$, wherein exact zero modes ($d_\fir(0) \!=\! 3$) are 
dimensionally different from lowest near-zero modes ($d_\fir(0^+) \!\simeq\! 2$). Few 
aspects of our results at $T \!=\! 234\,$MeV should be pointed out.
{\it (i)} The computed dimension $d_\fir$ of near-zero modes is in close vicinity of 
``topological value" 2, thus inviting a systematic inquiry into its possible origin in 
a certain topological feature of underlying glue fields. At the same time, recent 
findings of possible topological behavior in 3d Anderson model~\cite{Horvath:2022klk} 
also involve dimension $2$ but no glue fields. 
{\it (ii)} In the existing QCD data there is no clear evidence yet for critical value 
$d_\fir \!\approx\! 8/3$, which was suggested to be a generic feature of Anderson 
models~\cite{Horvath:2021zjk}.  
{\it (iii)} Unlike in the case of $\lambda_\fir$, we did not find an obvious dimension
jump in the vicinity of $\lambda_A$. This differs from the situation in pure-glue 
QCD~\cite{Alexandru:2021pap} and from that at critical points of Anderson 
models~\cite{Horvath:2021zjk}.
Taken together, points {\it (i-iii)} constitute an intriguing complex puzzle to be 
solved by future studies. 

In contrast to $T \!=\! 234\,$MeV, we found that the characteristic $\lambda \!=\!0$ 
discontinuity is absent at $T \!=\! 187\,$MeV. This is consistent with the original 
IR-phase proposal~\cite{Alexandru:2019gdm}, wherein there is a regime of rapid
but analytic change between the crossover temperature 
$T_\cro \approx T_c \approx 155\,$MeV and the onset $T_\fir$ of IR phase. 
[Note that $T_\cro$ denotes a generic crossover temperature characterizing rapid 
changes in Dirac spectra~\cite{Alexandru:2019gdm}.]
Nevertheless, as evidenced by Fig.~\ref{fig:d_IR_lambda}, there is a dimensional 
structure above the deep IR and below $T$, which is neither low-$d_\fir$ plateau
as in IR phase, but also not a constant $d_\fir \!=\! 3$ as at low temperatures. 
Resolving its nature will require better control over both IR and UV cutoff
systematics, which can be non-trivial in dynamical simulations~\cite{Zhao:2022ooq}
but achievable by current computational means.
Some aspects of cutoff effects are discussed in Appendix~\ref{app:uvcutoff}.

{The current approach to calculate $d_{\rm IR}$ is affected by finite volume 
effects, particularly in the plateau region of IR phase, and the associated 
"transition region" forming already below $T_\fir$ (see Fig.~\ref{fig:d_IR_lambda}). 
This complicates studies of dimensional properties and warrants testing the
alternatives such as the multidimension technique~\cite{Horvath:2022klk}}.

Recently a number of lattice QCD papers focused on the same temperature range as the one investigated 
here (see e.g.~\cite{Dick:2015twa,Ding:2020xlj,Aoki:2020noz,
Kaczmarek:2021ser,KOTOV2021136749, Chen:2022fid, Kaczmarek:2023bxb}). Their physics goals are mostly 
different and tend to involve the chiral limit, such as in studies of U$_\text{A}$(1) 
problem or chiral phase transition. 
From symmetries perspective, it would be interesting to study how low dimensional nature of 
low-lying Dirac modes in IR phase relates to the approximate color-spin symmetry observed in 
Refs.~\cite{Rohrhofer:2019qwq, Glozman:2022lda}. Other related 
developments can be found in Refs.~\cite{Liu:2023cse, Athenodorou:2022aay, Kehr:2023wrs}. 
The present CLQCD data could be used to study some of these problems.

\smallskip

\section*{Acknowledgments}
We thank the CLQCD collaborations for providing us their gauge configurations with 
dynamical fermions~\cite{Hu:2023jet}, which are generated on HPC Cluster of ITP-CAS, 
the Southern Nuclear Science Computing Center(SNSC) and the Siyuan-1 cluster supported by 
the Center for High Performance Computing at Shanghai Jiao Tong University. We also thank 
RBC collaborations for providing their gauge configurations, and Ting-Wai Chiu, Heng-Tong 
Ding and Jian Liang for valuable discussions. This work is supported in part by NSFC 
grants No. 12293060, 12293062, 12293065 and 12047503, a NSFC-DFG joint grant under Grant 
No.\ 12061131006 and SCHA 458/22, and also the Strategic Priority Research Program of 
Chinese Academy of Sciences, Grant No.\ XDB34030300 and YSBR-101. This work is also 
supported in part by the U.S. DOE Grant No.\ DE-SC0013065, DOE Grant No. DE-FG02-95ER40907, 
and DOE Grant No.\ DE-AC05-06OR23177 which is within the framework of the TMD Topical 
Collaboration. This publication received funding from the French National Research Agency 
under the contract ANR-20-CE31-0016. 
The numerical calculation were carried out on the ORISE Supercomputer through HIP 
programming model~\cite{Bi:2020wpt}, and HPC Cluster of ITP-CAS. This research used 
resources of the Oak Ridge Leadership Computing Facility at the Oak Ridge National Laboratory, 
which is supported by the Office of Science of the U.S. Department of Energy under Contract 
No.\ DE-AC05-00OR22725. This work used Stampede time under the Extreme Science and 
Engineering Discovery Environment (XSEDE), which is supported by National Science Foundation 
Grant No.\ ACI-1053575. We also thank the National Energy Research Scientific Computing Center 
(NERSC) for providing HPC resources that have contributed to the research results 
reported within this paper. We acknowledge the facilities of the USQCD Collaboration used 
for this research in part, which are funded by the Office of Science of the U.S. Department 
of Energy. 

\begin{appendix}
\begin{widetext}

\section{Accuracy of Overlap Eigenmodes}
\label{app:accuracy}

In this section we focus on the accuracy of the low-lying eigenvectors used in this study. 
For efficiency reasons, we compute the low-lying eigenvalues and eigenvectors of 
$M\equiv \frac{1}{\rho^2}D_{\rm ov}^{\dagger}D_{\rm ov} = 
\frac{1}{\rho}(D_{\rm ov}+D_{\rm ov}^{\dagger})$ 
in the chiral sector of exact zero mode(s). The eigenvalues and eigenvectors of $D$ 
are simply related to the eigenvectors of $M$: non-zero modes of $D_\text{ov}$ come in pairs, 
$\lambda_R\pm i\lambda_I$, and they span an invariant two-dimensional space for $M$ with 
$\lambda_M=\lambda_R$ and for near zero-modes $\lambda_R\approx \lambda_I^2/(2\rho)$. When 
computing the spectrum of $M$, zero modes appear as Ritz pairs with eigenvalues of the order 
of $\epsilon$, the precision of the sign-function approximation used to implement $D_\text{ov}$. 
When we have near-zero modes with $(a\lambda_I)^2 < \epsilon$, it is impossible to distinguish 
them from zero modes. Using a polynomial approximation for the sign functions, the best precision 
we are able to achieve is $\epsilon \approx 10^{-12}$, and consequently we can only confidently 
resolve eigenvalues with $a\lambda_I > 10^{-6}$, which in physical units correspond to 
$\lambda_I > 2\times 10^{-6}\,\text{MeV}$. For the volumes used in this study, the near-zero 
eigenmodes satisfy this condition.
 
Another concern is the mixing between nearly-degenerate eigenvectors. For eigenvector 
observables (like $f_\star$) that are smooth as a function of $\lambda$, this is less of 
a concern. However, at discontinuities mixing could introduce systematic effects. This could 
potentially be a problem at $\lambda=0$ since the zero modes and near-zero modes behave 
differently. We argue here that this is not the case.

To see this, consider two eigenvectors of the projected operator $D$ with
$D v_1=i\lambda_1 v_1$ and $D v_2=i\lambda_2 v_2$. A mixed vector 
$v=v_1 \cos\theta + v_2 \sin\theta$ has a Ritz ``eigenvalue'' $i\lambda=v^\dagger D v$ 
and residue $\delta \!=\! \Vert Dv - i\lambda v\Vert \!=\! 
|\cos\theta \sin\theta (\lambda_1 -\lambda_2)|$. 
The case relevant for us is $\lambda_1\approx 0$ and the near-zero value 
$\lambda_2 \!>\! 0.1\,$MeV where our residues, even in the worst case, 
are $\delta \!<\! 10^{-7}$. This implies that the mixing angle 
is at most $\theta\sim \delta/(\lambda_2a) < 2\times 10^{-3}$. Given that $f_\star$ 
varies slowly (the difference between zero modes and near-zero modes is less than 
two), this mixing will have negligible effect given our statistical errors.

\section{Fitting of Effective Fractions}
\label{app:fitting}

\begin{figure}[tbh]
\vspace{-0.12in}
\centering
\includegraphics[scale=0.435]{f_star_main.png}
\hspace{0.05in}
\includegraphics[scale=0.474]{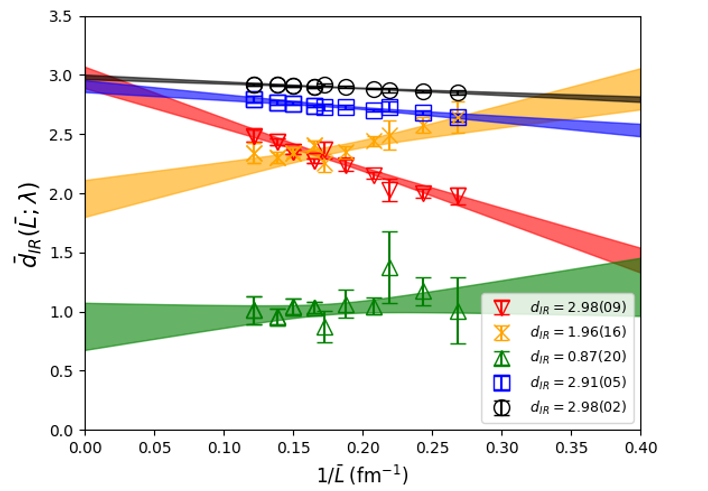}
\vspace{-0.08in}
\caption{Function $f_\star(L)$ (left) and the associated $\bar{d}_\fir(\bar{L})$ 
(right) for various $\lambda$ at $T\!=\!234\,$MeV. See text for explanations.} 
\label{fig:f_star_lambda_combine}
\end{figure}

Our procedure to extract $d_\fir$ assumes an approximately linear (in $1/L$) 
approach of ``finite-volume" dimension $d_\fir(L,s)$ in Eq.~(4) to its
$L \!\to\! \infty$ limit. This was suggested by Ref.~\cite{Horvath:2021zjk}
in the context of Anderson models. One can easily check that this is equivalent 
to direct fits of $\efNm(L)$ to the form $\efNm(L) \!\propto\! L^{d_\fir} e^{-c/L}$
and $f_\star(L) \!\propto\! L^{d_\fir -3} e^{-c/L}$ for effective volume fraction 
$f_\star(L) \!\equiv\! \efNm(L)/\nrN$. Fits shown in the left panel of 
Fig.~\ref{fig:f_star_lambda_combine} support the validity of this approach.

Here we wish to check the nature of finite-$L$ correction more directly. Given 
the above scaling form, $L$-dependence of IR dimension can be expressed as
\begin{equation}
   d_\fir(L,s) = d_\fir + \frac{s-1}{\ln(s)} \, \frac{c}{L} 
   \quad\;\; \longrightarrow \quad\;\;
   \bar{d}_\fir(\bar{L}) = d_\fir + \frac{c}{\bar{L}} 
   \quad\, \text{with} \quad\,
   \bar{d}_\fir(\bar{L}(L,s)) \equiv d_\fir(L,s) 
   \;\; , \;\;
   \bar{L}(L,s) \equiv L \,\frac{\ln(s)}{s-1}    \;\;
\end{equation}
 Introduction of variable $\bar{L}$ thus makes it possible to combine $d_\fir$ results 
 from all pairs of distinct lattices and follow their trends. Indeed, according to the above, 
 value $\bar{d}_\fir(1/\bar{L})$ from each pair should fall on an indicated straight 
 line, at least near $1/\bar{L} \!=\!0$. To check this, we show in the left plot of
 Fig.~\ref{fig:f_star_lambda_combine} our $f_\star$ data for five largest volumes
 at selected values of $\lambda$, and in the right 
 plot the associated functions $\bar{d}_\fir(1/\bar{L})$. Note that there are 10 data points 
 for each $\lambda$ in the latter case since this is the number of possible lattice pairs.  
 Displayed fits are indeed consistent with linear nature of $d_\fir(1/\bar{L})$ 
 near $1/\bar{L}\!=\!0$. The qualitative difference between exact zero modes and lowest
 near-zero modes is expressed in the right plot by the crossed lines representing
 $\lambda \!=\! 0$ and $\lambda \!=\!0.22\,$MeV. Their finite-volume corrections are 
 in fact of opposite sign.

\begin{figure}[h]
 \centering
 \hspace {0in}
 \includegraphics[scale=0.42]{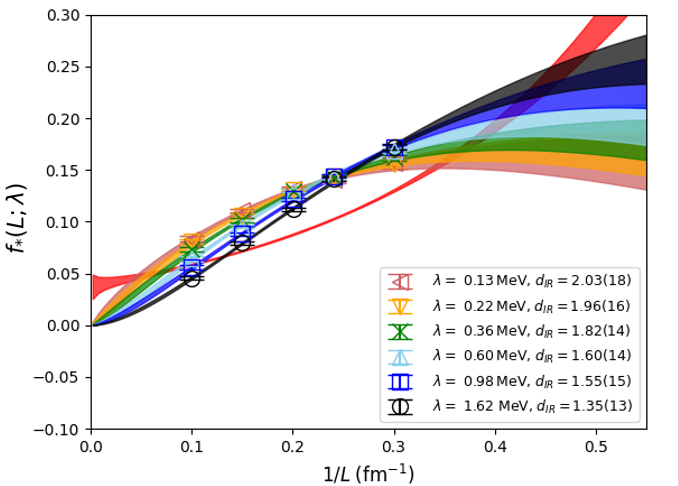}   
 \hspace{-0.15in}
 \includegraphics[scale=0.42]{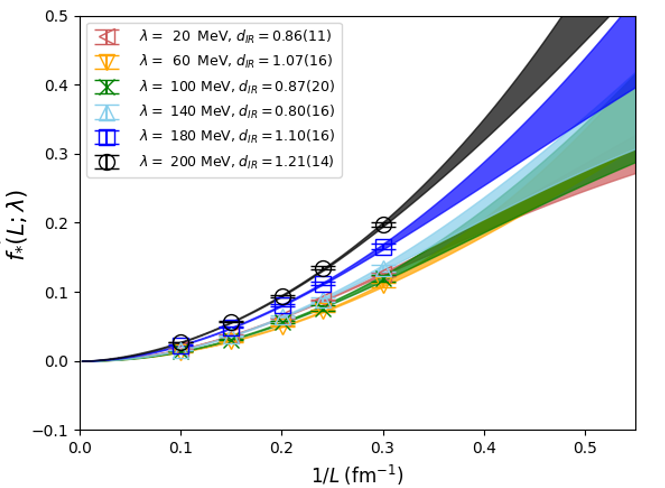}
 \hspace{-0.15in}
 \includegraphics[scale=0.42]{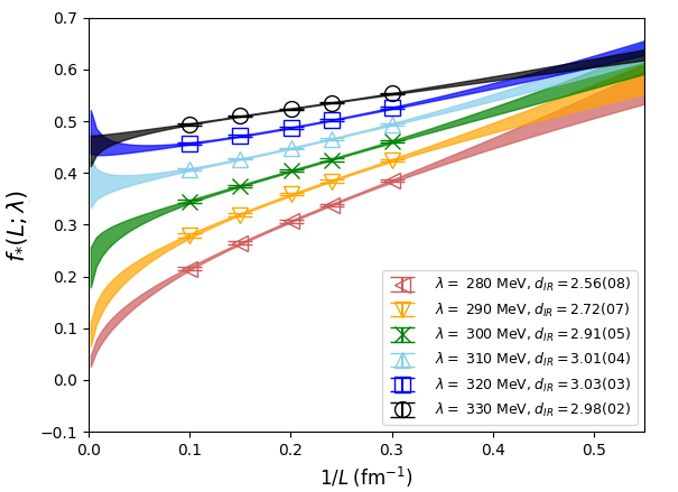}
 \caption{Function $f_\star(L)$ with different simulated size $L$ at $T$=234 MeV, 
 for $\lambda\in[0.1,2.0]$ MeV (left), $\lambda\in[20,200]$ MeV (middle) and 
 $\lambda\in[280,330]$ MeV (right).}
 \label{fig:f_star_lambda_cases}
\end{figure}

We also provide the $f_\star(1/L;\lambda)$ data for more values of $\lambda$ in 
Fig.~\ref{fig:f_star_lambda_cases}. As shown in the left panel, 
the $f_\star(L>4~\mathrm{fm};0<\lambda<2~\mathrm{MeV})$ becomes higher when $\lambda$ 
is smaller, and thus the corresponding $d_{\rm IR}$ is also larger. The tendency 
converges at $\lambda\sim$ 0.2 MeV (orange band), which corresponds to $d_{\rm IR}=2$ 
and is consistent with the $\lambda=$0.13 MeV case (dark red band) within 
the uncertainty. However, this limit is significantly different from $f_\star(;0)$ as 
illustrated by the red band. Thus, $f_\star$ and also $d_{\rm IR}$ will be discontinued 
at $\lambda=0$. 

In contrast, the middle panel of Fig.~\ref{fig:f_star_lambda_cases} shows that $d_\fir$ 
changes smoothly with $\lambda$ for $\lambda\in[20,200]\,$MeV, corresponding 
to $d_\fir \sim 1$ within $2\sigma$. The change of $d_\fir$ is also smooth 
in the range $\lambda\in[280,330]\,$MeV where it converges to 3 with increasing 
$\lambda$, as in the right panel.  Therefore the discontinuity of $d_\fir$ would 
only occur at $\lambda=0$, given our statistical precision at $a=0.105$ fm and 
$T=234\,$MeV.

\begin{figure}[h]
  \centering
  \includegraphics[scale=0.50]{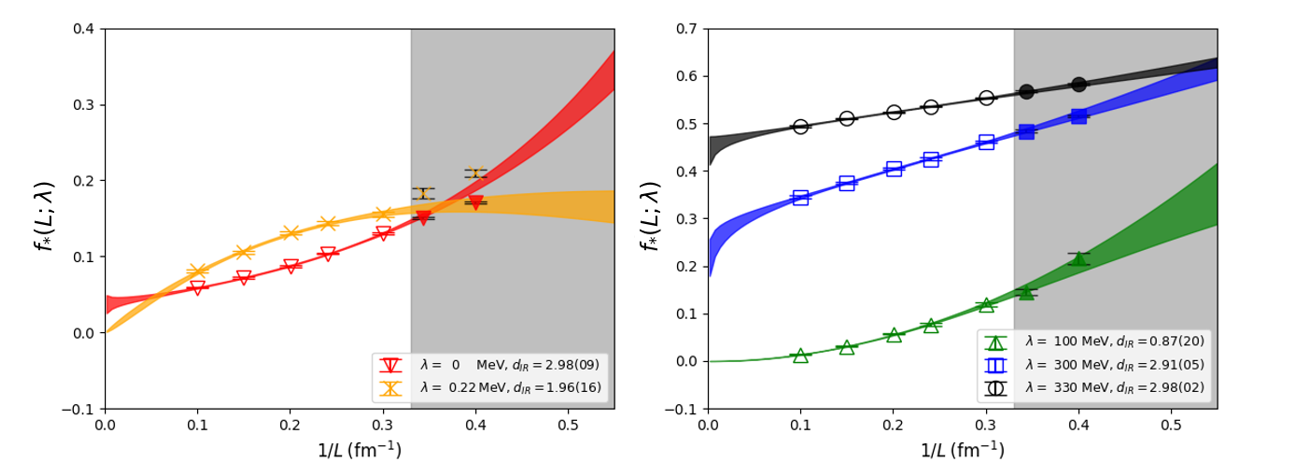}
  \caption{Function $f_\star(L)$ at $T\!=\!234\,$MeV in different spectral regions, for all 
  simulated sizes $L$. Left: zero modes and near-zero modes. Right: plateau, just below $\lambda_A$ 
  and at the bottom of the bulk. Shaded regions are excluded from the displayed fits.} 
  \label{fig:f_star_lambda_sm}
\end{figure}

Finally, we give a justification for using the five largest lattices in our fits i.e. systems 
with $L \!>\! 3\,$fm. To that end, we show in Fig.~\ref{fig:f_star_lambda_sm} functions 
$f_\star(1/L)$ for all simulated volumes, together with previously shown fits. Shaded areas mark 
the volumes excluded from these fits. One can see that in case of zero modes and near-zero modes 
(left plot), the systems in shaded region do not follow the fit curves, and were thus excluded 
from fits in all spectral regions. 

A similar plot of the results for $N_f \!=\! 2+1$ zero-temperature ensembles 
at $a=0.194\,$fm and $L \in \{5.7,6.2,9.3\}$ fm is shown in 
Fig.~\ref{fig:dist_lambda_ex_zero} (left panel). 
Note that $f_{\star}$ of exact zero modes is significantly different 
from that of non-zero modes, while $d_\fir$ is consistent with $3$ at all 
studied values of $\lambda$.

\begin{figure}[bth]
\vspace{-0.15in}
\centering
\includegraphics[scale=0.45]{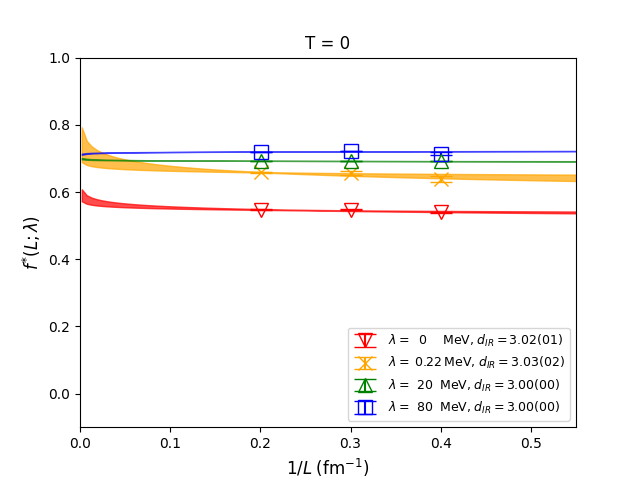}
\includegraphics[scale=0.6]{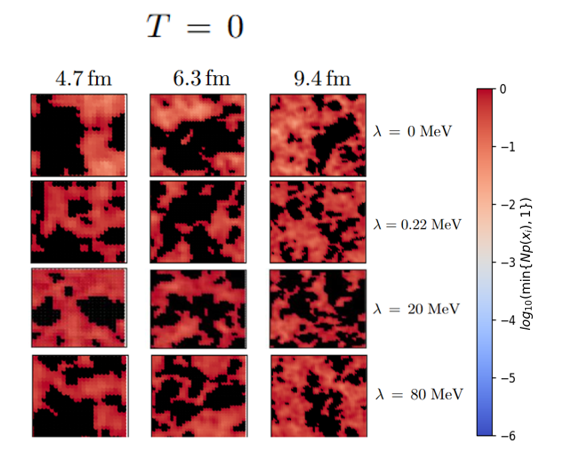}
\caption{
$f_{\star}(L)$ (left) and Typical color-coded $\log_{10}(\min\, \{\nrN p(x_i), 1 \})$ in a 2d plane 
containing point $x_i$ with maximal probability (right)
at zero temperature.} 
\label{fig:dist_lambda_ex_zero}
\end{figure}

\section{Visualization of Spatial Distributions}\label{app:Visualization}

In this Appendix we wish to extend our visualization of mode distributions in 
various $\lambda$-regimes at $T\!=\!234\,$MeV shown in Fig.~4 of the manuscript. 

In particular, the right panel of Fig.~\ref{fig:dist_lambda_ex_zero} demonstrates 
that zero-temperature mode distributions remain quite similar at different scales 
$\lambda$ and various IR cutoffs, and are qualitatively different from that shown in Fig. 4.

\begin{figure}[bth]
\vspace{-0.15in}
\centering
\includegraphics[scale=0.63]{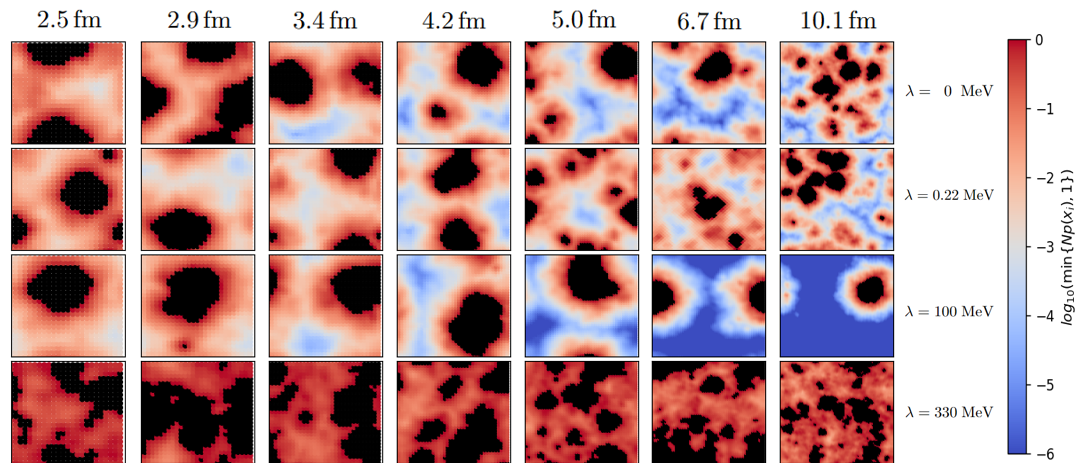}
\caption{
Typical color-coded $\log_{10}(\min\, \{\nrN p(x_i), 1 \})$ in a 2d plane 
containing point $x_i$ with maximal probability. Modes in different 
$\lambda$-regimes on a given glue background at $T\!=\!234$~MeV are shown 
for all spatial sizes $L$ studied.} 
\label{fig:dist_lambda_ex}
\end{figure}

In contrast, Fig.~\ref{fig:dist_lambda_ex} shows examples of modes at the same 
four values of $\lambda$, but at all IR cutoffs (sizes $L$ of the system) 
considered in this work. for $T$=234 MeV. Note that
$\lambda \!=\!\,$0, 0.22 and 100 MeV belong to the IR part of the spectrum, while 
$\lambda \!=\!\,$330 MeV is in a near-bottom part of the bulk. 

Interesting aspect of observing the typical geometry at fixed $\lambda$ for increasing 
$L$ is an evolution in degree of granularity. Indeed, note that for the plateau mode 
($\lambda \!=\! 100\,$MeV), increasing $L$ confirms the picture of a single solid lump 
present in the volume. On the other hand, for zero-modes and near-zero modes, the apparent
(visually observable) degree of granularity increases with increasing $L$, reflecting
that their effective supports keeps spreading out in the volume. In fact, all qualitative 
aspects we observe agree with {\em metal-to-critical} picture of transition to IR phase, 
put forward in Ref.~\cite{Alexandru:2021xoi}. The associated details will be worked out 
in dedicated follow-up publications. 

\section{Lattice Spacing Dependence}
\label{app:uvcutoff}

\begin{table}[t] 
\vspace{-0.10in}
\centering
\caption{UV cutoff $a$, pion mass $m_{\pi}$, lattice volumes $n_{L}^3\times n_T$ and temperature $T$ 
of lattice QCD ensembles studied.}
\label{Tab:setup_all}
\begin{tabular}{cclrc}
\hline
\hline
$a$(fm) & $m_{\pi}$(MeV) & $n_L$ & $n_T$ & $T$(MeV) \\
\hline
 0.105 & 135 & 24/28/32/40/48/64/96 & 8 & 234\\
&  & \quad \quad \ \ \  32/40/48/64 & 10 & 187\\
 0.105 & 290 & 24 & 8 & 234\\
\hline
0.052 & 317 & 48/ \quad \quad \quad \quad \! 96 & 16 & 234 \\
\vspace{-0.35in}
\end{tabular}
\end{table}

To assess the impact of lattice spacing on our findings, we recalculated $\rho(\lambda)$ 
and $f_{\star}$ at $a=0.052$ fm with a heavier pion mass of $m_{\pi}=317$ MeV, and also 
$\rho(\lambda)$ at  $a=0.105$ fm with $m_{\pi}=290$ MeV. Details regarding these ensembles 
are listed in Table~\ref{Tab:setup_all}.

\begin{figure}[tbh] 
\centering
\includegraphics[scale=0.5]{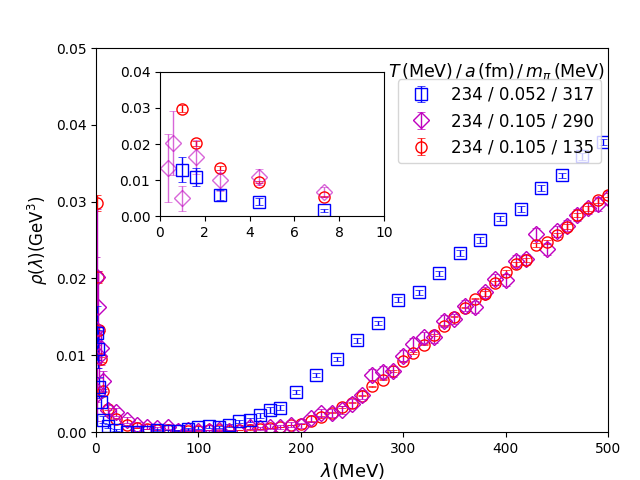}
\caption{Spectral density $\rho(\lambda)$ at $T \!=\!$ 234\, MeV at $a=0.105$ fm with 
$m_{\pi}$=135 MeV (red dots), $m_{\pi}=290$ MeV (tiny black boxes), and $a=0.052$ fm with 
$m_{\pi}$=317 MeV (blue crosses).}
\vspace{-0.21in}
\label{fig:densities_lat}
\end{figure}

In Fig.~\ref{fig:densities_lat}, pink diamonds show the $\rho(\lambda)$ at  $a=0.105\,$fm 
with $m_{\pi}=290$ MeV with relatively low statistics, and show good consistency {for $\lambda>$ 4 MeV} with that at 
the same lattice spacing but $m_{\pi}=135\,$MeV (red dots). It suggests that the pion mass 
dependence of $\rho(\lambda)$ {at relatively large $\lambda$} is weak, as shown in the previous study using the overlap fermion 
on the Domain wall sea~\cite{Aoki:2020noz}. Thus we calculate $\rho(\lambda)$ at smaller 
lattice spacing but heavier pion mass to investigate the lattice spacing dependence.

As shown as the blue crosses in Fig.~\ref{fig:densities_lat}, the peak of $\rho(\lambda)$ 
at small $\lambda$ diminishes at $a=0.052\,$fm with the same temperature but a heavier $m_{\pi}$. 
This decrease implies that the magnitude of this peak may contain a significant 
discretization error, such as the mixed action effect. At the same time, $\rho(\lambda)$ at 
$a=0.052$ fm becomes higher than that at $a=0.105$ fm for $\lambda>120$ MeV. This disparity 
is also likely primarily attributed to discretization errors.

\begin{figure*}
\vspace{-0.1in}
\centering
\includegraphics[scale=0.67]{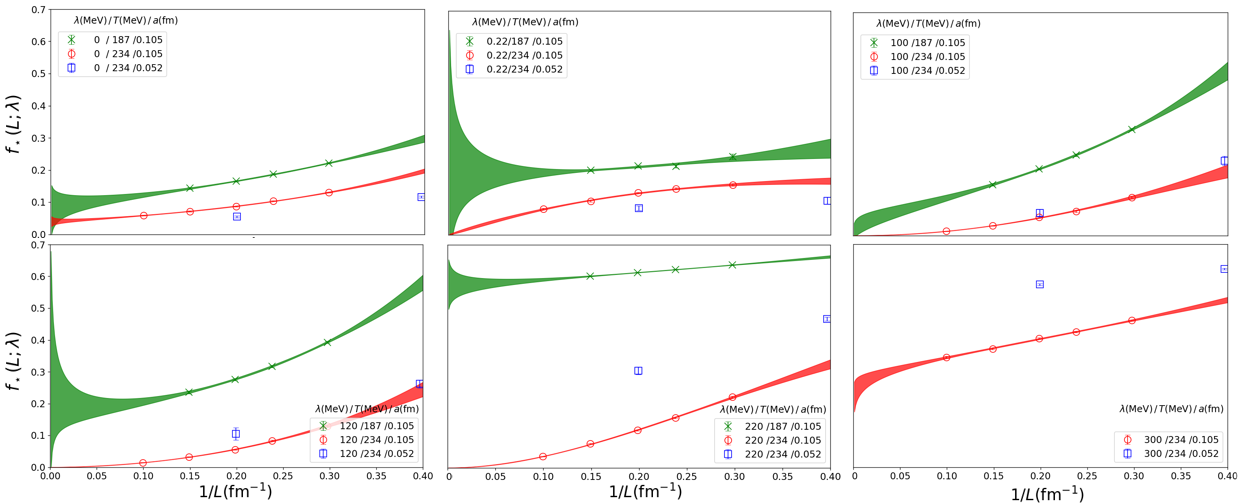}\\
\caption{Function $f_\star(1/L)$ for various values of $\lambda$, temperatures and UV cutoffs.} 
\label{fig:f_star_lambda_lat}
\end{figure*}

Regarding $f_\star$, we display the $f_\star$ values at two temperatures and $a=0.105\,$fm 
in Fig.~\ref{fig:f_star_lambda_lat} for various typical $\lambda$, and include the values at 
$T=234\,$MeV and $a=0.052\,$fm (blue boxes) for two volumes for comparison. The values at 
$a=0.052\,$fm are typically lower than those at $a=0.105\,$fm for $\lambda<100\,$MeV, 
indicating a sparse distribution in the infinite volume limit. However, $f_\star$ are larger 
for $\lambda>100\,$MeV, which can result in a decrease in $\lambda_\text{A}$ such that 
$d_{\fir}(\lambda>\lambda_\text{A})=3$ as we approach the continuum limit.

In addition, comparing the $f_\star$ with two temperatures at $a=0.105$ fm, those at 
higher temperature are always lower and suggest that the Dirac mode and then gluon field 
become ``sparser".

\end{widetext}

\end{appendix}

\bibliographystyle{apsrev4-1}
\bibliography{reference.bib}

\end{document}